\documentclass[sn-basic]{sn-jnl}

\jyear{2023}%

\theoremstyle{thmstyleone}%
\theoremstyle{thmstyletwo}%

\theoremstyle{thmstylethree}%

\usepackage{ulem}

\definecolor{thibautcolor}{HTML}{F37A60}

\usepackage{cleveref}
\usepackage{siunitx}
\usepackage{pdfpages}

\begin{document}

\title[Randomly stacked open shells]{Randomly stacked open-cylindrical shells as a functional mechanical device}


\author*[1]{\fnm{Tomohiko G.} \sur{Sano}}
\email{sano@mech.keio.ac.jp}
\equalcont{These authors contributed equally to this work.}
\author[2]{\fnm{Emile} \sur{Hohnadel}}
\equalcont{These authors contributed equally to this work.}
\author[1]{\fnm{Toshiyuki} \sur{Kawata}}
\author[2]{\fnm{Thibaut} \sur{M\'etivet}}
\author[2]{\fnm{Florence} \sur{Bertails-Descoubes}}



\affil[1]{\orgdiv{Department of Mechanical Engineering}, \orgname{Keio University}, \orgaddress{\city{Yokohama}, \postcode{2230061}, \state{Kanagawa}, \country{Japan}}}

\affil[2]{\orgdiv{INRIA, CNRS}, \orgname{Univ. Grenoble Alpes}, \orgaddress{\city{Grenoble}, \postcode{38000}, \state{INP, LJK}, \country{France}}}



\abstract{
Structures with artificial mechanical properties, often called mechanical metamaterials, exhibit divergent yet tunable performance. 
Various types of mechanical metamaterials have been proposed, which harness light or magnetic interactions, structural instabilities in slender or hollow structures, and contact friction. 
However, most of the designs are precisely engineered without any imperfections, in order to perform as programmed. 
Here, we study {the} mechanical performance of randomly stacked cylindrical-shells, which act as a disordered mechanical metamaterial. 
Combining experiments and simulations, we demonstrate that the stacked shells can absorb and store mechanical energy upon compression by exploiting large deformation and relocation of shells, snap-fits, and friction.
Although shells are oriented randomly, the system exhibits robust mechanical performance controlled by friction and geometry. 
Our results demonstrate that rearrangement of flexible components could yield versatile but predictive mechanical responses. 
}


\maketitle

\section{Introduction}\label{sec1}

Predicting the large deformation of slender structures, such as pillars, beams, and arches, is one of the central issues in material science~\citep{Gordon2003}.  
By forecasting structural instability and optimising their design to prevent rupture, {slender parts can be assembled into lasting architectural buildings, easier to build and able to withstand large external forces, thereby resilient to natural disasters}. 
Engineers {design} energy- or force-absorbing structures to protect humans and objects from impact or shocks. Examples include bicycle helmets and packaging materials for fragile baggage~\citep{LuYu2003}. The energy- or force-absorbing structures are often soft elasto-plastic multiscale materials, carefully designed at the structural and chemical levels, to control the deformation modes. 

When a slender beam is compressed axially, it bends and buckles, as bending is more energetically favourable than compression~\citep{Landau_book,Bazant_book}. 
Buckling of a slender beam can be regarded as a classical and canonical example of a -- reversible -- energy-absorbing phenomena~\citep{Reis2015}. The beam bends and buckles when subjected to forces beyond Euler's critical load, and
the external energy is then stored as bending energy. The stored energy can thereafter be released by removing the external load. 
Hence, the instability of slender structures could be interpreted as a mechanical energy transducer~\citep{Reis2015}. This idea has been part of the recent evolution of materials and structures with an artificial mechanical response, referred to as \textit{mechanical metamaterials}~\citep{Holmes2019}. 

Recent progress in fabrication technology and computational modelling have also fostered the investigation of tunable mechanical functionality. 
Various types of mechanical responses have been studied, such as auxetic materials~\citep{Lakes1987,Bertoldi2010}, friction-dominated assemblies~\citep{Poincloux2021}, snapping~\citep{Sano:2018ji,Fu2019,Yoshida2020}, twist-extension coupling~\citep{Frenzel2017}, elasto-magnetic coupling~\citep{Chen2021-rn}, or origamis and kirigamis~\citep{Silverberg2014}. 
{Mechanical metamaterials however usually require  accurate engineering to achieve the programmed responses~\citep{Bertoldi2017}. The effects of randomness and imperfection of the components on the mechanical performances are seldom clarified, although these factors are likely to manifest all the more as the diversity and local arrangement of the structures at play constantly increase.}

\begin{figure}[!h]
    \centering
    \includegraphics[width = 1.0\textwidth]{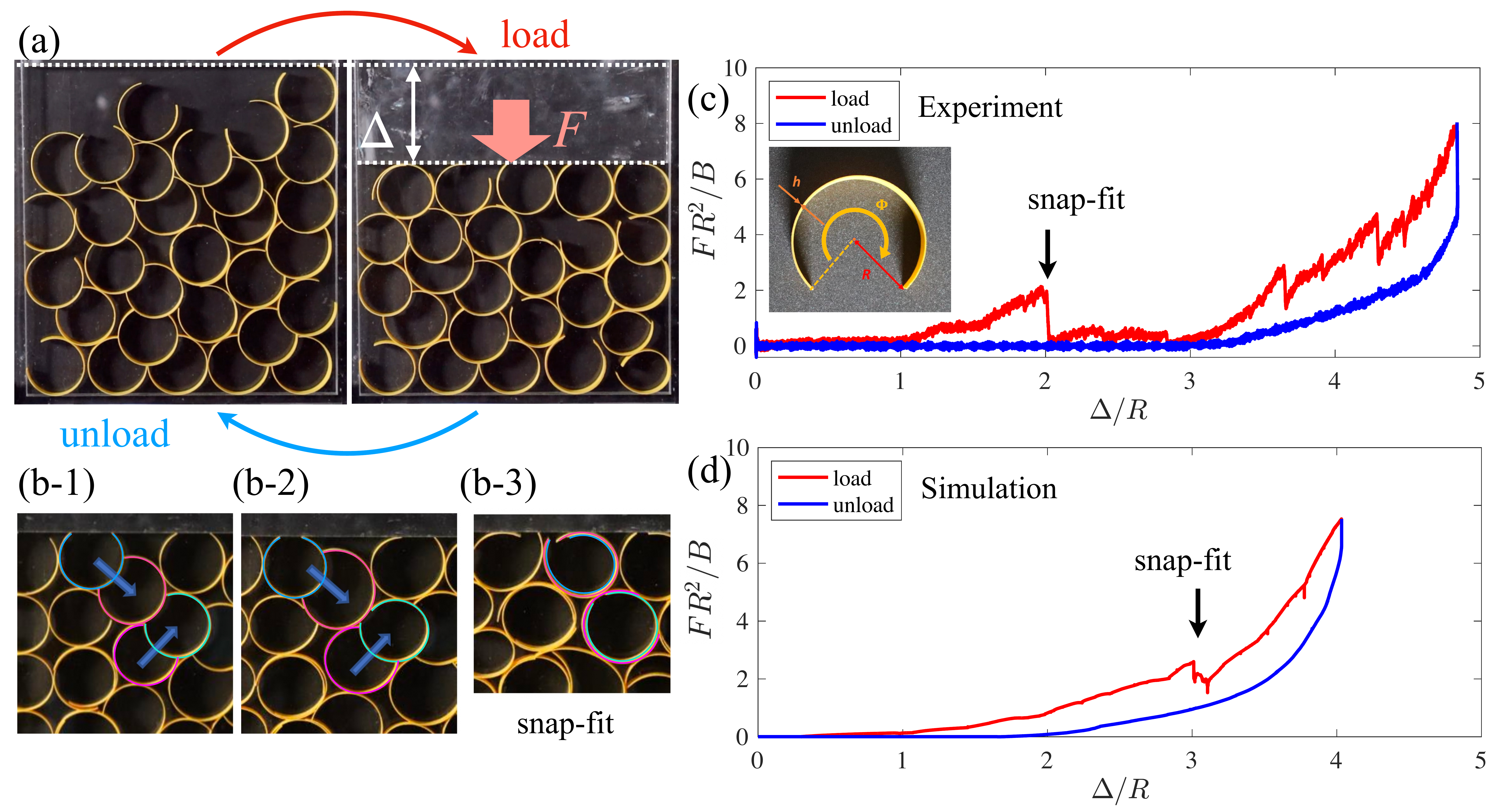}
    \caption{Mechanics of randomly stacked open-cylindrical shells with shell angle $\Phi$. (a)~{Our base experiment, where} the randomly stacked shells are compressed and then decompressed.
    (b-1)-(b-3) The load $F$ is absorbed into the elastic deformation with $F$ decreasing when shells embrace each other (snap-fit). 
    (c-d) The force-displacement curve upon compression (red) and decompression (blue) in (c) experiments and (d) simulations. The black arrows highlight the illustrative force-drops in the load $F$.}
    \label{fig:1}
\end{figure}

In this article, we report the mechanical properties of randomly-stacked open cylindrical shells, combining experiments and simulations(\cref{fig:1}(a)). {We carefully fabricate  open cylindrical thin shells whose rest geometry (see inset in \cref{fig:1}(c)) is an extruded surface characterised by a radius $R$ and a shell angle $\Phi$, the thickness $h$ being negligible compared to the shell length $\Phi R$. Each shell deforms elastically in 2D and interacts with other shells through contact and friction.}
We then consider a stack of identical shells placed in a random 2D configuration (see \cref{fig:1}(a)). This heterogeneous system exhibits characteristic damping against compressive loads, supplemented by friction, elastic bending, and reorientation of the shells. Upon compression, two shells interact with each other and may then ``overlap'' (a phenomenon called \textit{snap-fit}). These snap-fit events irreversibly reduce local empty spaces and voids, thereby lowering the overall compressive load. In addition, frictional contacts between the shells induce further dissipation throughout the cycle. 
As will be shown in the following, although the initial configuration of our system is set randomly,  energy dissipation remains nearly constant across statistical sampling, while compressibility turns out to be directly tunable by the shell geometry. Our work is complementary to the pioneering work by~\citet{Poirier:1992}, where localised deformation of stacked straws (closed cylindrical shells) is studied purely experimentally. 

\section{Results}\label{sec2}

\begin{figure}[!h]
    \centering
    \includegraphics[width = \textwidth]{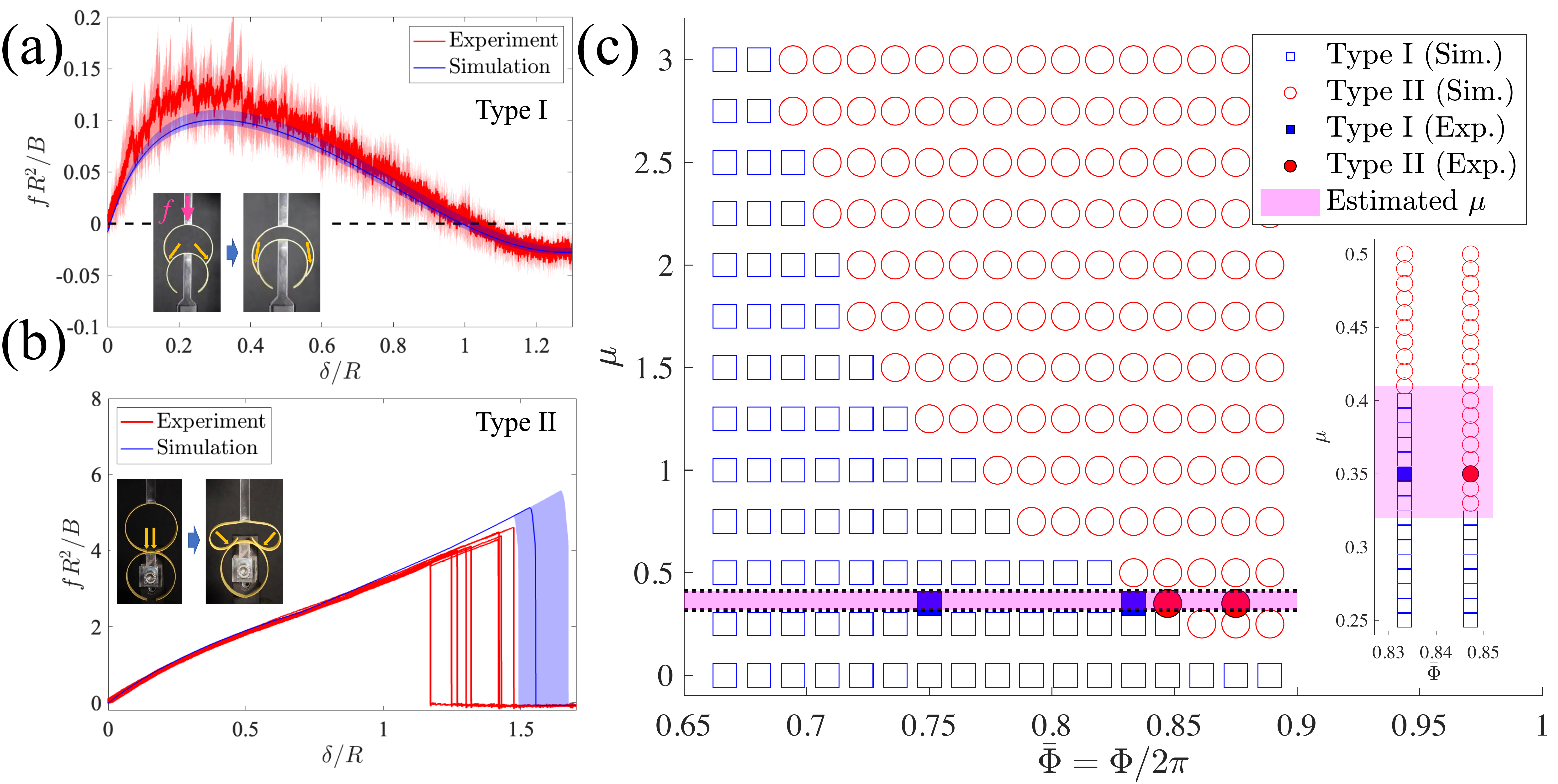}
    \caption{How an elastic shell snaps onto {a second identical shell}. (a-b) Rescaled force-displacement curves for (a)~Type~I and (b)~Type~II snapping regimes, where the red and blue curves are experimental and simulation results, respectively. (c)~Phase diagram of the shell deformation on the $(\Phi/2\pi)$--$\mu$ plane{, where empty data points are simulation results. In parallel, experiments made for various controlled open angles $\Phi$ \--- but for an uncertain friction coefficient $\mu$ \--- lead either to Type I or Type II outcomes (filled data points): this experimental capture of regime change allows us to identify precisely the range of the possible values for the experimental friction coefficient $\mu$ (shaded region) as the interval $[0.32 - 0.41]$, where the simulated Type~I and II phase boundary coincides with the experimental one. In our simulations we then systematically take $\mu = 0.35$ when comparing to experimental data.}}
    \label{fig:2}
\end{figure}

Identical randomly stacked open cylindrical shells are compressed and decompressed in a force-testing machine. In the absence of any external load ($F=0$), empty spaces still exist between the shells, which gives rise to a porous structure, contributed by both the friction between shells and their elasticity~(\cref{fig:1}(a)). Upon compression, the load $F$ increases with large deformation of the shells, followed by force drops corresponding to (local) intermittent snap-fits. As we compress the system further, the increase and decrease of the compression force continue up to the end of the test~(\cref{fig:1}(b)-(d)).

{\paragraph{Elementary two-shell snap-fit}}
The observed drops of the total force of the system find their origin in local snap-fit phenomena {involving} two open shells, where one shell embraces the other upon pushing. To understand the elementary process {responsible for the} mechanical performance of the stacked shells, we consider a compression test between two identical shells with varying shell angle $\Phi$. 
When two identical shells are assembled, they snap either smoothly or abruptly, depending on the friction coefficient $\mu$ and the shell angle $\Phi$. We call the former and the latter modes Type I and II snap-fits respectively, following the classification that~\citet{Yoshida2020} {proposed in the case of a shell indenting against a rigid cylinder}.
{In the Type I snap-fit, the upper indenting shell continuously slides on the bottom one, passing the smooth peak force $f_{\rm Type I} \sim 0.1 B/R^2$ (with $B$ the bending modulus of the shell), until the indentation exceeds the radius of the shell, when the force becomes negative and the shells naturally snap-fit each other. The Type II behaviour, which can be observed for more closed shells, on the contrary features sticking and coiling of the upper shell, until some threshold indenting force is reached, and the shell abruptly unfolds onto the other. Note that the Type II threshold force is much higher -- $f_{\rm Type II} \sim 5B/R^2\gg f_{\rm Type I}$ -- than the typical forces involved during Type I events.}
In \citet{Yoshida2020}, the shell-cylinder problem is studied with different diameter ratios, leading to either snap-fit or misfit, when the shell cannot fit onto the cylinder. However, given that we consider two identical shells, our shells always snap-fit, and we shall only consider Type I and II snap-fits throughout.

We investigate the two-shell elementary process both experimentally and numerically (see \cref{fig:2}).
It is noteworthy that due to its 2D nature, the two-shell assembly scenario can be simulated using 2D thin elastic rods, by taking the Poisson ratio into account in the re-scaling of the Young modulus. To this end, we develop an unclamped, 2D version of the {\it Super-Helix} model for discrete Kirchhoff rods~\citep{BACQLL06} with a full account of dry frictional contact~\citep{DBB11}. In \cref{subsec:simulations} we provide a detailed description of our numerical rod simulator coupled to frictional contact, which has been recently validated in both 2D and 3D configurations~\citep{RMRCLNB21}{, in particular on the ``pinning test"~\citep{SYW17} which couples elasticity and frictional contact}. {We run a new validation test, by comparing our simulator to the analytical master curve of \citet{Yoshida2020} in the case where the shell and the cylinder feature the same radius (see fig. 1 in Supplementary Information)}. The excellent agreement observed here again between our 2D rod simulations and our shell experiments for both the shape geometry and the force-displacement curves in Type I and II regimes confirms the validation of our numerical setup for naturally curved rods {and for a large range of friction coefficients (the pinning test being in contrast limited to the $[0,0.33]$ range}), and paves the way to extensive parametric studies.

Using exhaustive simulations, we explore the ``phase'' diagram of the shell-shell snap-fit mechanism, and find{, similarly to the shell-cylinder analysis of ~\citet{Yoshida2020},} that both the friction coefficient $\mu$ and the shell geometry $\Phi$ impact the snapping behaviour between two identical shells. As summarised in \cref{fig:2}(c), the larger the friction coefficient $\mu$, the more Type II snap-fit happens for open-angle shells. Given the friction coefficient~$\mu$, there exists a critical shell angle $\Phi^* = \Phi^*(\mu)$ between the Type I and II regimes. {Experimentally, we similarly find out a transition between Type I and Type II regimes, as $\Phi$ is increased.
These experimental and numerical results highlight the fact that not only the friction coefficient between shells but also their geometric non-linearity plays a critical role in the mechanical performance of stacked shells.}
{Furthermore, it is noteworthy that} thanks to its monotonous dependence, the critical angle $\Phi^*(\mu)$ can be exploited to determine accurately the friction coefficient between our experimental shells by measuring the actual threshold angle between the Type~I and Type~II behaviours, as explained in \cref{fig:2}.

In the remainder of the paper, we thus use this experimental measurement to bound the friction coefficient $0.32\leq\mu\leq0.41$ and use $\mu=0.35$ in all our simulations (two-shell and many-shell) for comparison with the experiments. Note that this value is consistent with the independent measurement we obtain using the pinning test protocol on a straight shell~\citep{SYW17, RMRCLNB21}{, for which we observe no sliding, meaning that $\mu > 0.33$.} 
{With this $\mu$ value fixed, we thus classify our experimental shells into two categories depending on their normalised angle $\bar{\Phi}\equiv\Phi/(2\pi)$: Type I shells for $\bar{\Phi}\leq 0.83$, and Type II shells for $\bar{\Phi} > 0.83$.}

{\paragraph{Many-shell snap-fit}}

\begin{figure}[!h]
    \centering
    \includegraphics[width = 0.8\textwidth]{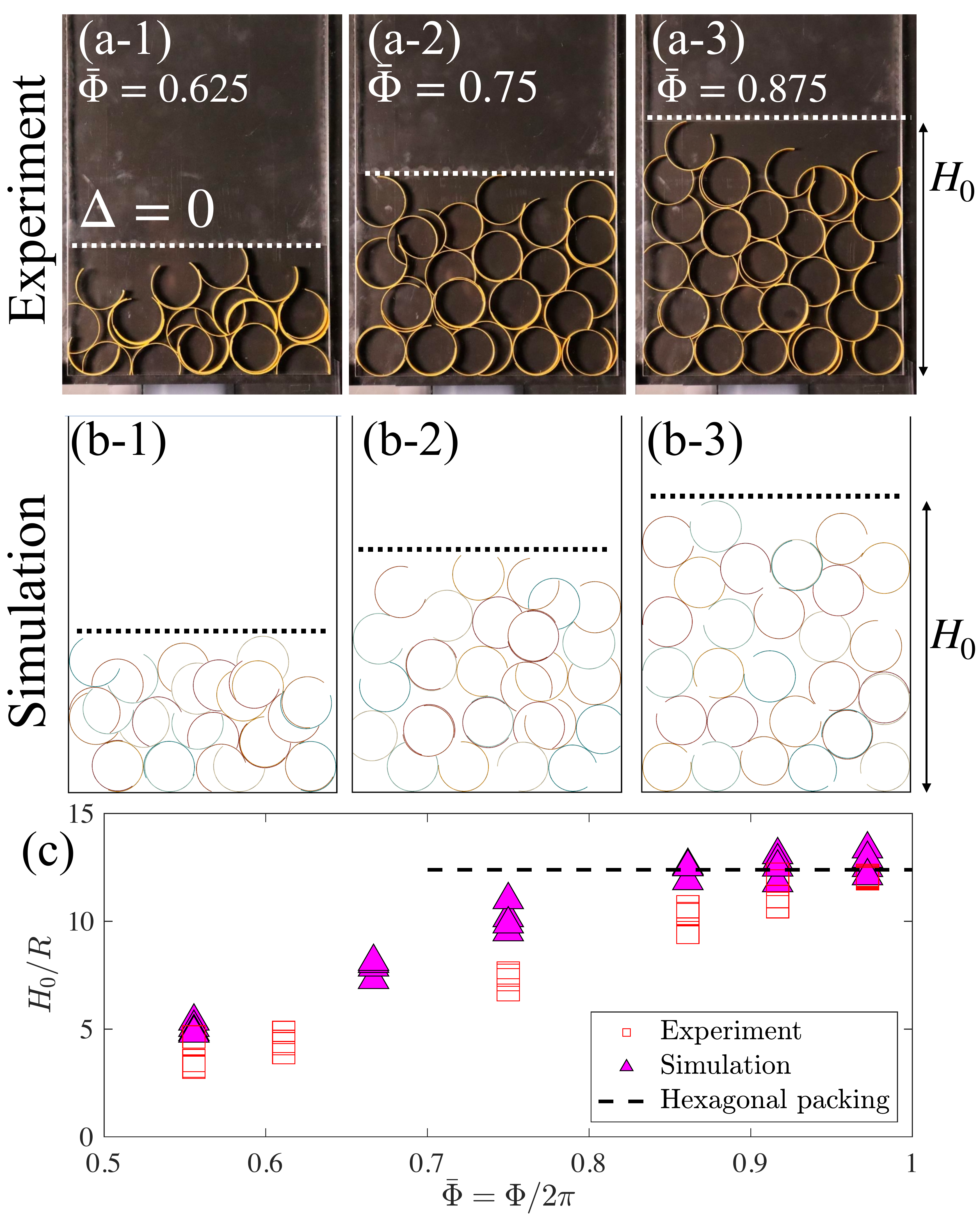}
    \caption{Static properties of randomly stacked shells under gravity (before compression). Configurations of shells obtained from (a-1)-(a-3)~experiments and (b-1)-(b-3)~simulations with various $\Phi$. (c)~Normalised height (prior to compression), $H_0/R$, as a function of $\Phi$.}
    \label{fig:3}
\end{figure}

We have systematically fabricated sets of $N=30$ shells with various $\Phi$ ({all other parameters remaining fixed}) and stacked them randomly inside a container {under the effect of gravity} (c.f.~\cref{fig:3}(a) and (b) for typical static configurations). It is noteworthy that the geometry of shells controls the height of the full system when stacked. Indeed, the initial height of the stacked shells, $H_0$, monotonically increases with $\Phi$ and saturates to the height corresponding to the hexagonal packing $H_0/R\to2+6\sqrt{3}$ as $\Phi \to 2\pi$. We observe that the simulation correctly captures the experimental and expected behaviour for $H_0$.

\begin{figure}[!h]
    \centering
    \includegraphics[width = \textwidth]{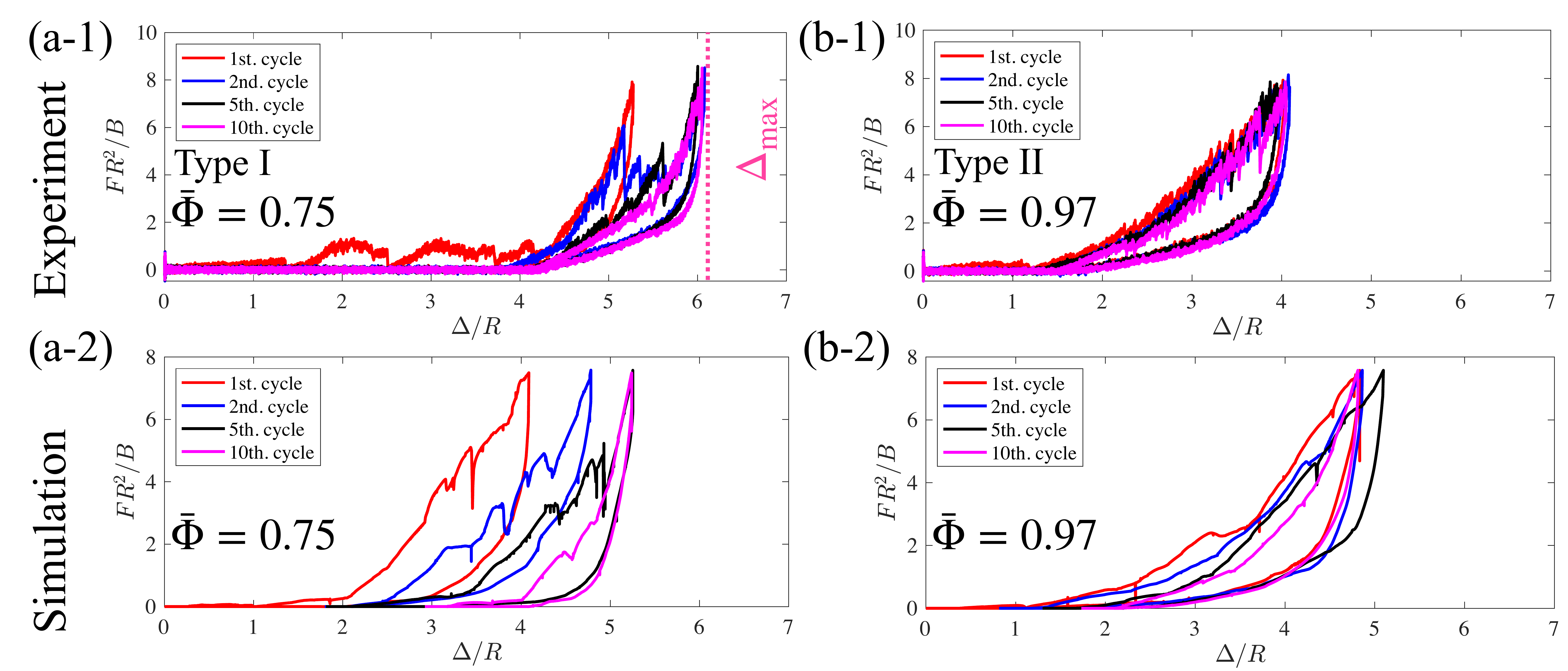}
    \caption{Dynamic properties of {30} randomly-stacked shells. Force-displacement curve for Type~I~($\bar{\Phi}=0.75$)  (a-1)(a-2) and Type~II shells~($\bar{\Phi}=0.97$) (b-1)(b-2). As the number of cycles increases, the curve converges to a limit cycle loop{, that is reached much more quickly for Type II shells}.}
    \label{fig:4}
\end{figure}

To further analyse the mechanical properties of the stacked shells, we perform cyclic compression-decompression experiments. The stacked shells are compressed from $\Delta \equiv H_0 - H = 0$ by a rigid plate, up to the point where the force exerted on the plate reaches some maximal value $F=F_{\rm max}$. The material is then decompressed back to $\Delta=0$~{(see \cref{fig:4}(a-b))}. 
In practice, we use $F_{\rm max} = 5~{\rm N}$ throughout the experiments and simulations, to yield as many snap-fits as possible while ensuring that we remain much below the plasticity and rupture thresholds of the shells. 

When we start compressing the stacked Type I shells {($\bar{\Phi}\leq 0.83$)}, in the first $n=1$ cycle the compressive load exhibits a characteristic damping behaviour: the force $F$ increases as the shells bend and then drops abruptly as snap-fit events occur within the system~(\cref{fig:4}(a-1),(a-2)). 
The continued compression causes snap-fits throughout the system. A pair of overlapped shells could be regarded as a shell of double thickness $2h$, thus increasing (effective) bending stiffness and stiffening the overall system. As a result, $F$ increases rapidly as we compress. Upon decompression, the assembled shells often do not separate into two, so that $F$ gets relaxed to zero smoothly. The compression and decompression process is {as such} fully irreversible. For the second $n=2$ {and subsequent} cycles, snap-fit is less and less likely to happen, because the force required for the Type~I snap-fit $f_{\rm Type I} \approx 0.06~{\rm N}$ is much lower than $F_{\rm max}$: $F_{\rm max}\gg f_{\rm Type I}$. In other words, most of the possible ``snappable'' pairs have already snapped after the 1st compression. 
As the cycle continues $n\gg1$, the force-displacement curve converges to a limit cycle curve, where elastic bending of the shells and frictional sliding between them are dominant. 

In contrast, when we compress the stacked Type II shells {($\bar{\Phi} > 0.83$)}, the force-displacement curve is qualitatively different from that of Type I shells. Interestingly, we do not observe Type II snap-fit in this problem set, despite that the compressive force is higher than the Type II threshold: $f_{\rm Type II}\sim 5B/R^2 \simeq 3.13~{\rm N} < F_{\rm max}$. As we observe in the two-shell snap-fit, the Type~II shell sticks, rolls and then unfolds upon compression, requiring much space for its large deformation. However, surrounding shells prevent opening the shell. In other words, the compression induces elastic bending and sliding only. As a result, the limit force-displacement cycle curve is basically immediately reached. This qualitative difference in the mechanical performance of stacked shells originates not only from the elasticity and geometry of each shell but also from contact mechanics and rearrangement of shells.

In both Type I and Type II cases, the limit force-displacement cycle exhibits a characteristic dissipation hysteresis {which is reminiscent of other thin elastic systems coupled with frictional contact~\citep{andrade2021cohesion}.} 
To understand quantitatively the dissipative properties of this system, we compute the maximum working distance, $\Delta_{\rm max}$, and the dissipation ratio $\eta$ (defined below) {in our $N=30$ shells experiments and simulations}. {Note that to ensure the robustness of the results and mitigate the effect of the initial conditions for such small-size systems, we assume that the system is ergodic and perform several independent measurements with respectively $5$ and $100$ different random initial configurations for the experiments and simulations. In addition, we use the robust normalised mean absolute deviation to estimate the standard deviation, as it decreases the bias of potential outliers~\citep{huber2011robust}.} In the following, we restrict the analysis to the first $n=1$ cycle, where the snap-fit effects are more pronounced and characteristic than the limit cycles $(n\geq 2)$.

\begin{figure}[!h]
    \centering
    \includegraphics[width = \textwidth]{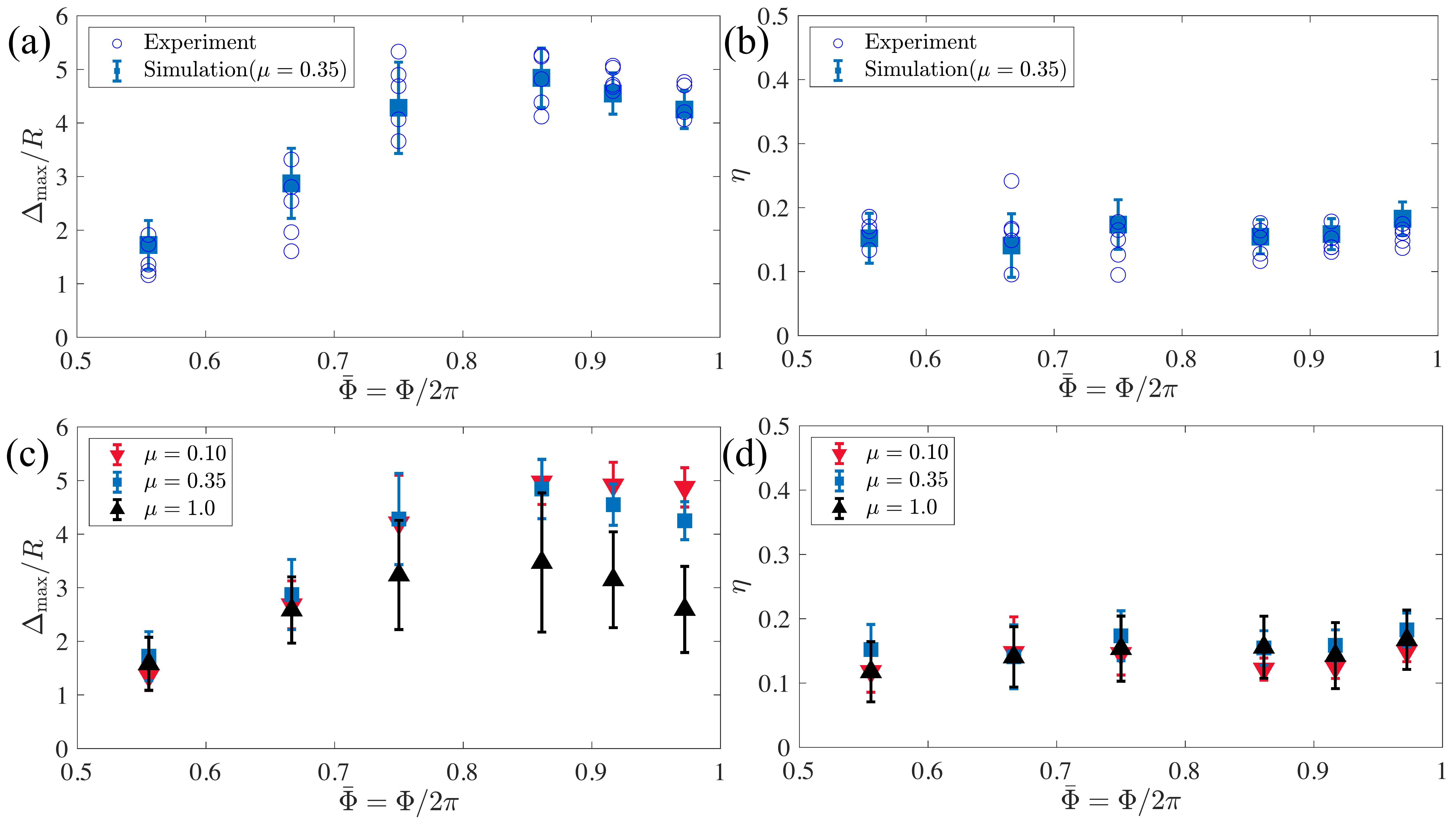}
    \caption{The maximum relative displacement, $\Delta_{\rm max}$ and energy absorption ratio, $\eta$ as a function of $(\Phi,\mu)$. Experimental and simulation results are superposed. Experiment results for (a)~$\Delta_{\rm max}$ and (b)~$\eta$, plotted against $\Phi$ as open symbols. The corresponding simulation results for different friction coefficients ($\mu=0.35$) are plotted as filled symbols. Simulation results of (c)~$\Delta_{\rm max}$ and (d)~$\eta$ for different friction coefficients ($\mu=0.10,0.35,1.0$).}
    \label{fig:5}
\end{figure}

The maximum working distance $\Delta_{\rm max}$ represents the displacement between the initial height $H_0$ and the final height of the cycle, when $F$ reaches the prescribed maximum force $F_{\rm max}$. 
In the Type I regime, snap-fit events lead to force drops and therefore increase the final total displacement required to reach $F_{\rm max}$. The more closed the shells, the larger the force drops, so that $\Delta_{\rm max}$ increases with $\Phi$~(\cref{fig:5}(a)). The increase of $\Delta_{\rm max}$ as a function of $\Phi$ continues up to the critical angle $\Phi^*(\mu)$ between the Type I and II regimes. For Type II shells, $\Delta_{\rm max}$ is nearly constant, as shells bend and slide without snapping upon compression, so that the force-displacement relation is mostly driven by shell elasticity. 
The ability to tune $\Delta_{\rm max}$ with shell geometry $\Phi$ could be useful in order to program the compression properties on demand. 

In addition to $\Delta_{\rm max}$, we also measure the dissipation ratio, $\eta$, defined as the area of the hysteretic $F$-$\Delta$ curve, $E_{\rm diss}\equiv\oint F d\Delta$, normalised by the total energy input $E_{\rm in}\equiv F_{\rm max}\Delta_{\rm max}$: $\eta \equiv E_{\rm diss}/E_{\rm in}$. 
Despite the qualitative differences observed between small-$\Phi$ and large-$\Phi$ $F$-$\Delta$ curves, the dissipation ratio $\eta$ turns up to be independent on the opening angle $\Phi$. We can therefore conclude that the stacked assembly exhibits robust energy-dissipating performance.

Note that the results from the simulations at $\mu=0.35$ are in excellent agreement with the experimental data for $(\Delta_{\rm max},\eta)$, as shown in \cref{fig:5}(a) and (b). Given the prior validation of our numerical method on shell-on-shell experiments, we shall thus fully rely on numerical simulations to study the role of the friction coefficient $\mu$.

To evaluate the role of friction in the mechanical and dissipative performances of the system, we carry out numerical simulations for smaller $(\mu=0.10)$ and larger $(\mu=1.0)$ friction coefficients, using $N=30$ shells and $100$ different initial random configurations as before. The results for $(\Delta_{\rm max},\eta)$ are summarised in~\cref{fig:5}(a) and (b), respectively. 
While $\Delta_{\rm max}$ seems independent of $\mu$ for shallow ($\bar{\Phi} \lesssim 0.75$) shells, it decreases as $\mu$ increases for deeper shells: the transition angle $\Phi^*(\mu)$ being a decreasing function of $\mu$~(c.f. \cref{fig:2}), the larger the friction coefficient, the more the system becomes Type~II dominated, with less snap-fit events -- and therefore less force drops -- occurring for $F \lesssim F_{\rm max}$.

In contrast, $\eta$ appears constant over the wide range of $\mu$, indicating robust performances in terms of dissipation properties. This is counter-intuitive because snap-fits seem irrelevant to the dissipation of the stacked shells. The shell geometry $\Phi$ and friction coefficient $\mu$ control $\Delta_{\rm max}$ as well as the total dissipation $E_{\rm diss}$ by the same amount, which would lead to the nearly robust~$\eta$.
Given that the value of $\eta$ results from cooperative effects between shells, additional theoretical effort will be necessary to predict $\eta$, which we leave as future work.

Based on the simulation results, we can therefore conclude the following design principle for stacked shells: the combination of $\mu$ and $\Phi$ determines the maximum working distance, $\Delta_{\rm max}$, of the system. Given the value of $\Delta_{\rm max}$, we can require the set of $\mu$ and $\Phi$, keeping the dissipation ratio $\eta$ nearly constant.

\section{Discussion}\label{sec3}

Our system exhibits a characteristic energy-absorbing behaviour, which could be exploited as a functional damper. The dissipation mechanism behind the stacked shells stems from the elasticity and geometry of shells, their contact mechanics, and their relative orientations. In this paper, we have studied the working distance and dissipation ratio $(\Delta_{\rm max},\eta)$ for the first ($n=1$) compression-decompression process. 
Through $n=1$ compression, stacked shells become stiffer, because snap-fitted shells are much stiffer than a single shell (by doubling thickness as $\sim 2h$ effectively). After the initial press, most of the possible shells have already snapped, leading the force-displacement curve to the limit cycle for $n\geq 2$. 
Even though shells do not snap so often at large $n$ cycles, stacked shells are still flexible (through elastic bending). 
Given that conventional shock-absorbing materials exploit the collapse or fracture of their elements (voids, foams), the idea of stacking slender structures is useful in designing functional structures.

The mechanical metamaterials studied in the previous literature exhibit artificial mechanical performances upon compression, which primarily rely on programmed precise elementary structures of the systems. The presence of defects and imperfections are, as such, critical in their mechanical performances~\citep{Evans2015}.
{The initial random orientations of the shells play, in our case, the role of such structural defects, and while the exact evolution of the system during compression is dependent on the initial conditions, the overall mechanical performances have been found to be nearly robust.}
Hence, large deformation of slender structures can be utilised even when they are stacked randomly. Our study paves the way for new designing principles, where slender structures deform largely and relocate their positions from each other. We expect that a similar design idea will be applicable to engineering problems across different length scales from food packaging to the deformation of bowl-shaped molecules~\citep{Furukawa2021}. 

In order to directly compare our numerical and experimental results while keeping computational times down, we have performed simulations for $N=30$ shells. However, our simulation protocol allows to robustly simulate significantly larger systems, albeit at the expense of high computational burdens: systems composed of $N\sim10^3$ shells for instance require, on a standard PC, about $9$ days of computation per compression-decompression cycle, mainly owing to the very accurate resolution of frictional contact constraints required to prevent penetrations and crossings of the shells.  
Obtaining reliable average and error bars for $(\Delta_{\rm max},\eta)$ therefore requires much more computational power for very large systems, and we leave the study of the thermodynamic limit of our system ($N\gg1$) as future work (c.f. \cref{sec:statistics}).

A fully analytical model would also provide a valuable and complementary approach towards the analysis of such large stochastic slender assemblies.
Approaches based on statistical mechanics concepts have proven successful in predicting the physics of random systems, such as for example molecular gases~\citep{Landau_book_kin}, granular particles~\citep{BP2004} or active matter systems~\citep{Kanazawa2020}. Kinetic theory (e.g. Boltzmann equation) provides an efficient framework to predict the phase dynamics for these systems. However, the analytical approaches so far have been limited to rigid materials or simple contact configurations~\citep{Poincloux2021}, and the large deformation of the composing structures are not taken into account in the present kinetic theory. Our experimental results will therefore be valuable to validate extended kinetic theories and models for largely-deformable components. 

\section{Methods}

\subsection{Fabrication of open cylindrical shells}\label{exp:fab}

Open cylindrical shells are casted in a thermoplastic way from naturally straight ribbons made of Glycol-modified Poly Ethylene Terephthalate (PET-G). The straight ribbon is laser-cut into a length~$L$ and width $w=10$~mm, from a flat sheet (Young's modulus, $E=2.2~{\rm GPa}$, thickness $h=0.5$~mm, Shim Stock, Artus Corporation, USA). The total length of the ribbon, $L$, is varied in the range of $66.3\leq L\leq116.1$~mm, such that the shell-angle $\Phi (=L/R)$ ranges $200^{\circ}\leq\Phi\leq350^{\circ}${, meaning that $0.55 \leq \bar{\Phi} \leq 0.97$}.
The surface of the straight ribbon is roughened by sandpaper to reduce adhesive forces.

The mould for the open shells consists of two parts; outer and inner moulds. The acrylic outer mould has a circular hole, while the inner circular mould, which fits into the hole of the outer mould, is made of silicone elastomer (Smooth-on, USA). The straight ribbon is inserted between the outer and inner moulds such that the ribbon is bent into the uniform radius of curvature $19$~mm. The set of a ribbon and mould is placed into the hot water of temperature $95\mathrm{^\circ C}$ for more than 15 minutes. Subsequently, the ribbon and moulds are cooled in cold water ($\sim20\mathrm{^\circ C}$) and then demoulded. The radius of curvature of the shell becomes $R = 19.2\pm0.2$~mm.

\subsection{Experimental protocols for two-body problems}\label{exp:two}

Two identical shells are compressed with each other quasi-statically at the speed of $0.17{\rm mm/s}$. The middle of the shells is either fixed or connected to the force-testing machine (EZ-LX, Shimadzu, Japan). The former shell is fixed with the ground via an acrylic bar of $4$~mm width. The latter shell is glued with an acrylic bar, which is clamped with the load cell.

\subsection{Experimental protocols for randomly-stacked shells}\label{exp:random}

To prepare the initially stacked configuration, we randomly drop 30 shells one by one into the acrylic quasi-two-dimensional container ($200\times300\times11$~mm) by free-fall. 
The mechanical tests start at the initial height of the stacked shell {under gravity}, from which the zero of the displacement, $\Delta=0$, is set. 
The shells are compressed and decompressed by an acrylic plate of $8$~mm width with the speed of $5{\rm mm/s}$. The shells are compressed up to $5$~N and then decompressed back to $\Delta=0$. We iterate the compression/decompression cycles 10 times, against 5 different initial configurations. After completing each 10$^\textrm{th}$ cyclic test, we carefully examine the geometry of all the shells whether they are damaged or not. If damaged shells are found, we apply the same protocol as \cref{exp:fab} prior to the next mechanical test.

\subsection{Simulations}
\label{subsec:simulations}

\paragraph{2D thin shells as 2D Kirchhoff rods}
We represent our elastic shells in two dimensions using the two-dimensional Kirchhoff thin elastic rod model, which accounts for linear bending elasticity and exact geometrical non-linearities, at the origin of the large displacements of the rod~\citep{AP10}. To solve for the dynamics of this model, we develop a 2D, unclamped version of the high-order \textit{Super-Helix} model, which has been originally popularised in Computer Graphics in the context of hair simulation~\citep{BACQLL06}. This model can be seen as a Galerkin (weak) discretisation of the Kirchhoff equations, composed of~$N$ helical elements with uniform material curvatures and twists. In 2D, the degrees of freedom of the rod boils down to~$N$ scalar curvatures, and each element takes the form of a circular arc. Compared to a nodal model, such a curvature-based model presents several advantages: the automatic capture of inextensibility, linear bending forces that can be integrated implicitly without additional cost, and a better speed of convergence with the number $N$ of elements. In all the simulations performed in this paper, we took~$N=15$ elements for each rod, as we found this resolution sufficient for our accuracy needs. 

Like the continuous 2D Kirchhoff rod model, the 2D Super-Helix model (namely the \textit{Super-Circle} model) is parametrised by only four physical parameters: its length~$L$, its natural curvature~$\kappa^0$, its linear mass~$\lambda$ and its bending modulus $B$. When subject to gravity~$g$, the model can be characterised by two dimensionless parameters: the gravitational bending parameter~$\Gamma = {\lambda g L^3}/{B}$ and the curliness~$c = \kappa^0 L$.

\paragraph{Dry frictional contact}
To couple two elastic shells together, we model dry frictional contact through the Signorini-Coulomb law, which poses constraints on the admissible velocity and force at contact so that non-penetration and Coulomb friction conditions are satisfied. For the sake of simplicity, we only consider a single friction parameter~$\mu$, hence making no distinction between static and dynamic friction parameters. In our scenarios, we found that this single coefficient law was sufficient to yield very good agreements between experiments and simulations. We discretise the full, non-smooth dynamic problem of~$n$ contacting rods by using the Moreau time-stepping method~\citep{Moreau94}, which resolves frictional contact constraints implicitly.  In practice, we use the \href{so-bogus}{https://gitlab.inria.fr/elan-public-code/so-bogus} library which offers a free, robust, and efficient implementation of non-smooth frictional contact solvers based upon the hybrid algorithm first proposed by~\citet{DBB11}. We note that the coupling between the Super-Circle model and so-bogus has been carefully validated~\citep{RMRCLNB21} on the stick-slip scenario of a straight indenting rod (``pinning test"), first introduced by~\citet{SYW17}. We further improve the model by devising an arc-arc detection scheme, and validate our complete numerical model on a curved rod contacting a cylinder~\citep{Yoshida2020} (see Supplementary Information, section~I). For all our results we used a time-step of $10^{-4}$ s, a solver tolerance of $10^{-16}\, (\textrm{N}.\textrm{s})^2$ (square impulses) and a maximum number of iterations of $1000$, allowing the solver to reach an accuracy of $10^{-9}\,(\textrm{N}.\textrm{s})^2$ on average at each time-step.

The excellent agreements we obtain allow us, on the one hand, to calibrate precisely the frictional coefficient of our experimental shells by relying on simulation, using a shell-shell pinning experiment (see \cref{fig:2}), and on the other hand, to compare successfully our 30-shell experiment to simulation results (see \cref{fig:3,fig:4}). Finally, our confidence in the numerical simulator allows us to conduct extensive parametric studies on the many-body scenario, in which we vary both the shell angle~$\Phi$ and the friction coefficient $\mu$, and use a set of 100 different initial conditions per simulation for robust statistical output data (see \cref{fig:5}, next section and Supplementary Information, section~II). We conducted these simulations using the same protocol as the experimental one, in particular we used the speed of $5{\rm mm/s}$ of the upper plate, which guarantees a quasi-static regime.

\subsection{Statistics of $\Delta_{\rm max}$ and $\eta$}\label{sec:statistics}
We observe experimentally and numerically that the $\Delta_{\rm max}$ and $\eta$ quantities measured on our $30$-shell scenario are highly dependent on the initial configuration of the shell stack, which leads to some significant spreading in our results. A solution to reduce this spreading  would be to scale up drastically the number of shells, both experimentally and numerically. Relying on a $\sqrt{N}$ law for the reduction of the distribution variance, we however anticipate that more than $\num{10 000}$ shells would be necessary to yield less than $1\%$ of spreading. Considering that our experimental shells are fabricated manually, and that our simulation time raises from a few minutes for the first compression cycle of $30$ shells to a few days for that of $\num{1 000}$ shells, this makes this approach currently intractable, both from an experimental and numerical point of view.

Instead, for the sake of efficiency we stick to our small-scale $30$-shell scenario and, for each pair $(\Phi, \mu$), we run in parallel 100 simulations featuring each a different random initial configuration. This allows us to build robust statistical data for $\Delta_{\rm max}$ and $\eta$ across the variation of initial conditions. {Our plots are summarised in Supplementary Information}, for each $(\Phi, \mu)$ pair considered. They all feature a Gaussian distribution, which confirms the ergodicity assumption of our system. Moreover, it is noteworthy that the variances are small, which allows us to report reasonably accurate averages for both quantities (see fig. 2 in Supplementary Information). For the experimental values $(\mu = 0.35)$, it is noteworthy that our error bars are close to the experimental ones (see \cref{fig:5}).


\backmatter

\section*{Declarations}

\bmhead{Funding} This work was supported by MEXT KAKENHI 18K13519, JST FOREST Program, Grant Number JPMJFR212W (T.G.S.). E. H. is funded by a Ph.D. grant from ENS Lyon.

\bmhead{Conflict of interest} The authors declare no competing interests.

\bmhead{Availability of data and materials} Experimental and simulation data are available upon request. 

\bmhead{Authors' contributions} T.G.S., E.H., T.K., T.M., and F.B.-D. designed the research and interpreted the results. T.G.S. and T.K. performed experiments. F.B.-D. designed and implemented the unclamped super-helix model and E. H. and T.M. devised the circular arc detection algorithm. E.H. implemented the virtual 2D shell experiments and performed the numerical simulations, and T.M. and F.B.-D. supervised the numerical investigations. T.G.S managed the project. T.G.S., T.M., and F.B.-D. wrote the paper.

\bibliography{sn-bibliography}




\section*{Supplementary material}

# Supplementary Information: Randomly stacked open-cylindrical shells as a functional mechanical device

Tomohiko G. Sano,[1, *] Emile Hohnadel,[2] Toshiyuki Kawata,[1] Thibaut Métivet,[2] and Florence Bertails-Descoubes[2]

[1]*Department of Mechanical Engineering, Keio University, Yokohama, Kanagawa, 2230061, Japan*
[2]*INRIA, CNRS, Univ. Grenoble Alpes, Grenoble, 38000, INP, LJK, France*

## I. VALIDATION OF THE 2D *SUPER-HELIX* MODEL COUPLED TO *SO-BOGUS* AGAINST THE SHELL-CYLINDER PROBLEM

The simulation code used throughout our paper consists of the 2D *Super-Helix* rod model [2] coupled to the non-smooth frictional contact solver of Daviet *et al.* [3] implemented in the so-bogus library. In the main text of our paper, we study, as an elementary micro-structural event, the problem of snap-fitting between two open cylindrical shells, where one shell is compressed against the other one. To validate our code beforehand, we use a slightly different scenario consisting of the snap-fit of a cylindrical shell onto a rigid cylinder of same radius, referring ourselves to the fully-validated theoretical framework of Yoshida and Wada [1]. Note that the latter work is more general, as it considers the case where the radius $R$ of the shell can be different from that of the cylinder, $R_c$, with $\alpha \equiv R_c/R$ the ratio between the two radii. Using the condition of mechanical equilibrium and the Amontons-Coulomb law with a friction coefficient $\mu$, the analytical formula for the phase boundary between Type I and II is derived as [1, Supplemental, eq. S18]

$$\frac{K(\phi)}{K_\perp(\phi)} + \frac{\tan \varphi_0(\phi, \alpha) - \mu}{1 + \mu \tan \varphi_0(\phi, \alpha)} = 0, \qquad (1)$$

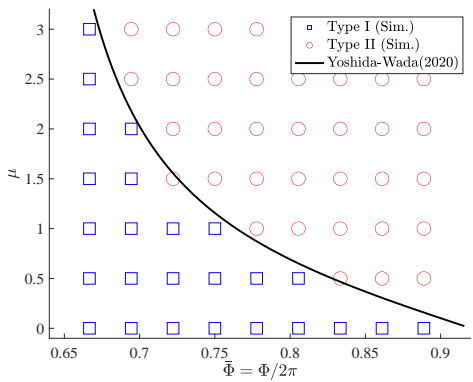

FIG. 1. The phase diagram of Type I and II snap-fits for the shell-cylinder problem. Empty squares and circles are Type I and II snap-fits, respectively. The solid black curve represents the theoretical curve, which is validated experimentally in Yoshida and Wada [1].

---

[*] sano@mech.keio.ac.jp

where $K(\phi), K(\phi)$ and $\varphi_0(\phi,\alpha)$ are known functions:

$$K(\phi) \equiv \left[\frac{\phi}{2} - \cos\phi\left(\frac{3}{2}\sin\phi - \phi\cos\phi\right)\right]^{-1} \quad (2)$$

$$K_\perp(\phi) \equiv \left[\frac{1}{2} - \cos\phi\left(-1 + \phi\sin\phi + \frac{3}{2}\cos\phi\right)\right]^{-1} \quad (3)$$

$$\varphi_0(\phi,\alpha) \equiv \sin^{-1}(\alpha^{-1}\sin\phi). \quad (4)$$

Here, we have introduced the half of $\Phi$ as $\phi \equiv \Phi/2$ (note that Yoshida and Wada [1] use $\Phi$ as our half angle). To validate our numerical model, we run simulations for different $\Phi$ and $\mu$ against this shell-cylinder problem, while the radii ratio is fixed throughout as $\alpha = R_c/R = 1$. In this case, equation (1) predicts a single-valued master curve on the $\Phi$-$\mu$ plane.

We summarise our simulation results superposed on the analytical curve of Yoshida and Wada [1] in Fig. 1. Our simulation results are in excellent agreement with equation (1). Given that equation (1) is validated experimentally in Yoshida and Wada [1], we can also conclude that the 2D *Super-Helix* model coupled to *So-Bogus* will be able to predict physical experimental results.

We stress here that this validation is important because it enriches previous validations made on our numerical model [4]. Indeed, in the latter reference, this code has been validated against the "pinning test" [5], where an elastic strip is vertically compressed against a rigid flat substrate. However, the validation protocol based on the pinning test is limited to the case when the strip only contacts the substrate punctually, which limits the validation range of $\mu$ to $0 \leq \mu \lesssim 0.33$. In other words, with this test, the simulation model is not validated for large $\mu$. In contrast, the snap-fit-based protocol in Yoshida and Wada [1] enables us to test the cases of large friction coefficients. Hence, we can conclude that the predictability of our numerical simulator applies to a large range of friction coefficients.

## II. HISTOGRAM FOR $\Delta_{\max}$ AND $\eta$ AGAINST DIFFERENT $(\bar{\Phi}, \mu)$

In the main text, we characterise our 30-shell experiment by defining the two descriptors $\Delta_{\max}$ and $\eta$. Thanks to the computational efficiency of the 2D *Super-Helix* model coupled to *So-Bogus*, we could obtain sufficient statistics for $\Delta_{\max}$ and $\eta$ across a range of $6 \times 3$ values for $(\bar{\Phi}, \mu)$, by using 100 different initial conditions per tuple $(\bar{\Phi}, \mu)$. All the data, gathered in fig. 2 ($\Delta_{\max}$ in green and $\eta$ in blue), feature Gaussian distributions with small standard or median absolute deviations.

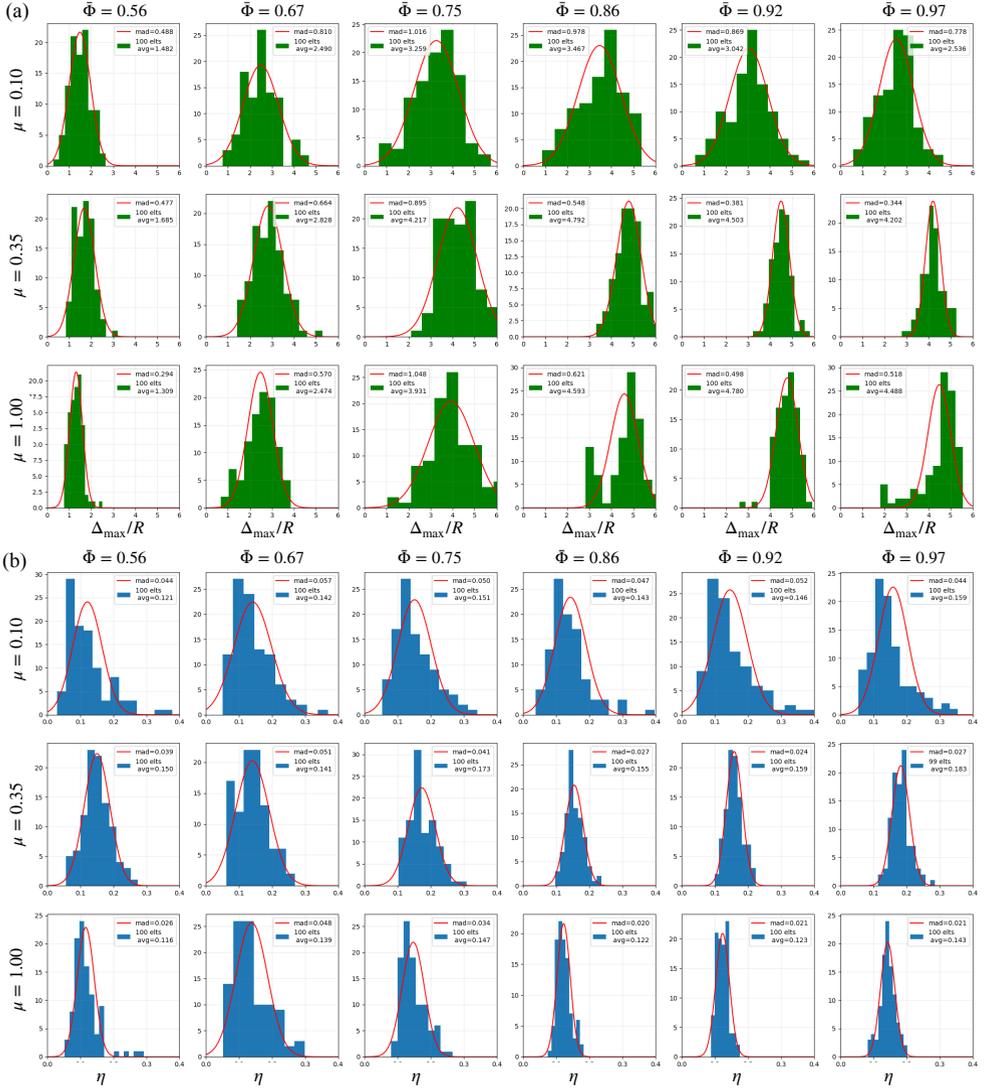

FIG. 2. Histogram of (a) $\Delta_{\max}$ and (b) $\eta$ for 100 different initial conditions. Solid lines are Gaussian distribution functions with the average (avg) and median absolute deviation (mad) calculated from the statistics of each $(\bar{\Phi}, \mu)$ simulation data.



\end{document}